\documentclass[twocolumn,preprintnumbers,superscriptaddress]{revtex4}
\usepackage{amsmath}
\usepackage{amssymb}
\usepackage{amsmath}
\usepackage{epsfig}
\usepackage{graphicx}
\usepackage{inputenc}
\inputencoding{latin1}
\usepackage{ulem}

\begin{document}

\title{Discovering baryon-number violating neutralino decays at the LHC}

\author{Jonathan~M.~Butterworth}
\affiliation{Department of Physics \& Astronomy, University College
London, UK}
\author{John~R.~Ellis}
\affiliation{Theory Division, Physics Department, CERN, CH-1211 Geneva 23, 
Switzerland}
\author{Are~R.~Raklev}
\affiliation{DAMTP, University of Cambridge, Cambridge CB3 0WA, UK;\\
Cavendish Laboratory, University of Cambridge, Cambridge CB3 0HE, UK}
\author{Gavin~P.~Salam}
\affiliation{LPTHE, UPMC Univ.\ Paris 6, CNRS UMR 7589,
Paris, France}

\begin{abstract}
Recently there has been much interest in the use of single-jet mass
and jet substructure to identify boosted particles decaying
hadronically at the LHC. We develop these ideas to address the
challenging case of a neutralino decaying to three quarks in models
with baryonic violation of R-parity. These decays have previously been
found to be swamped by QCD backgrounds. We demonstrate for the first
time that such a decay might be observed directly at the LHC with high
significance, by exploiting characteristics of the scales at which its
composite jet breaks up into subjets.
\end{abstract}

\preprint{CAVENDISH-HEP-2009-07, CERN-PH-TH/2009-073, DAMTP-2009-40}

\maketitle

The LHC's scope for the discovery of supersymmetry broken at the TeV
scale~\cite{Wess:1974tw,Fayet:1976cr,Dimopoulos:1981zb,Nilles:1983ge,
Haber:1984rc} has generated much interest. Certainly, the potential
prize is great: TeV-scale supersymmetry could solve several puzzling
problems and answer a number of open questions in modern particle
physics, such as the fine-tuning of the Higgs mass, the unification of
forces at high energies and the nature of dark matter. Effort has
mainly been concentrated on investigations into the discovery reach
and possibility of parameter measurements in the Minimal
Supersymmetric Standard Model (MSSM) and various more constrained
versions featuring a weakly-interacting and stable Lightest
Supersymmetric Particle (LSP), which gives a missing-energy
signature. Candidates for the LSP include the lightest neutralino,
$\tilde\chi_1^0$, and the gravitino.

However, the gauge symmetries of the Standard Model (SM) also allow
for dimension-four terms in the superpotential of the forms
\begin{equation}
\lambda_{ijk} L_iL_j\bar E_k + \lambda^\prime_{ijk} L_iQ_j\bar D_k
+ \lambda^{\prime\prime}_{ijk} \bar U_i\bar D_j\bar D_k,
\nonumber
\end{equation}
which violate the R-parity that is imposed in the MSSM. Non-zero values
for the couplings $\lambda$ could imply drastically different
phenomenologies for supersymmetry at the LHC, allowing for the decay
of the LSP, or any heavier sparticle, directly to SM particles.

Broadly speaking, we can classify the RPV models in terms of the
dominant coupling $\lambda$ and the identity of the Next-to-Lightest
Supersymmetric Particle (NLSP), assuming that a gravitino is the
LSP. The lepton-number-violating (LV) couplings $\lambda$ and
$\lambda^\prime$ generally give MSSM-like signatures with missing
energy from neutrinos and/or extra leptons in the decays of the
NLSP. The exception is a slepton/sneutrino NLSP that decays into two
quarks via a dominant $\lambda^\prime$ coupling. Scenarios such as
these should be easy to extract from the SM backgrounds, as shown for
a neutralino NLSP in~\cite{:1999fr}.

However, dominant baryon-number-violating (BV) couplings
$\lambda^{\prime\prime}$ are more difficult to deal with, due to the
large hadronic activity expected at the LHC, which threatens to drown
decays such as $\tilde\chi_1^0\to qqq$. The QCD background for jets
with $p_T>500$~GeV and a $\tilde\chi_1^0$ mass of
$\mathcal{O}(100~{\rm GeV})$ is about two orders of magnitude higher
than the signal. The background's non-trivial shape means that it is
hard to establish whether a small deviation from the expected
background is a signal of something new, or simply a defect in one's
understanding of the background. Some success has been reported by
relying on the production of a high-$p_T$ lepton in the decay chain
leading to the NLSP~\cite{:1999fr,Allanach:2001xz,Allanach:2001if},
but ideally one would wish to demonstrate the feasibility of signal
isolation and mass measurement in a less model-dependent
manner. Otherwise, the fear is that supersymmetry could escape
discovery at the LHC by cloaking itself in BV decays.

In this Letter we investigate that problem by looking for jets from
the decays of very boosted sparticles via BV couplings. Such decays
give rise to composite jets made up of two or more collimated
subjets, with a jet mass related to that of the original sparticle,
with specific properties predicted for the scale at which the main jet
separates into subjets. Similar techniques have previously been used
by the authors for analysing $WW$
scattering~\cite{Butterworth:2002tt,Aad:2009wy}, for detecting massive
boson decays in MSSM scenarios~\cite{Butterworth:2007ke} and in Higgs
searches~\cite{Butterworth:2008iy}. In addition, a number of other
techniques for separating hadronic decays of heavy particles from QCD
backgrounds have been suggested by other
groups~\cite{Agashe:2006hk,Brooijmans:2008se,
Kaplan:2008ie,Thaler:2008ju,Almeida:2008yp,Krohn:2009zg,Ellis:2009su}.
While some of these studies are applicable to general searches, they
all worked with examples of particles with known or well-constrained
mass. The present article tackles a new problem: how to identify a
hadronic resonance of unknown mass in an explicitly scale-invariant
manner.
%
The techniques presented here in a supersymmetric
scenario with RPV clearly have applications to any
hadronically-decaying massive-particle resonance that can be produced
far above threshold, and are promising for broad use in the
challenging LHC searches for hadronic decays of new particles.


We focus our investigation on the CMSSM benchmark point SPS1a.
We also look at six other CMSSM points with larger sparticle masses
and lower cross sections, that lie along the corresponding benchmark
line~\cite{Allanach:2002nj}. These encompass sparticle masses from
$(m_{\tilde\chi_1^0},m_{\tilde q},m_{\tilde g})\simeq
(96,530,600)$~GeV to $(161,815,915)$~GeV, and cross sections from 47
pb to 4.6 pb at leading order. We note that the gluino and squark
masses are much larger than the neutralino mass,
and can hence yield a highly-boosted neutralino in their decays.

R-parity violation is incorporated by setting the dominant coupling to
be $\lambda^{\prime\prime}_{112}=0.001$. This coupling is chosen to be
difficult: no heavy flavours are present to help tag the correct jets
and the coupling is chosen to be relatively large, so that decays do
not lead to displaced vertices \cite{constraints}.

In order to simulate sparticle pair-production events at the LHC with
RPV decays, we use the {\sc Herwig~6.510} Monte Carlo event
generator~\cite{Corcella:2000bw,Dreiner:1999qz,Moretti:2002eu,Corcella:2002jc}
with {\sc CTEQ 6L}~\cite{Pumplin:2002vw} PDFs, and use {\sc
Jimmy~4.31}~\cite{Butterworth:1996zw} for the simulation of multiple
interactions \cite{jimmy}.  This is interfaced to the {\sc
FastJet~2.4.0}~\cite{Cacciari:2005hq,FastJetWeb} jet-finder package
using the {\sc Rivet}~\cite{Waugh:2006ip} framework. Our background
sample, consisting of QCD $n$-jet events, $t\bar t$, $W$+jet, $Z$+jet
and $WW/WZ/ZZ$ production is simulated with the same setup, with the
exception of the $n$-jet events where we use {\sc Alpgen
2.13}~\cite{Mangano:2002ea} interfaced to {\sc Herwig/Jimmy} to
generate up to 4-jet final states with the MLM jet-parton matching
scheme. For the QCD event generation we also require $\sum
p_T>600$~GeV to reduce the large sample size, where the sum is over
all jets. The leading-logarithmic parton shower approximation that is
used has been shown to model jet substructure well in a wide variety
of processes~\cite{Abazov:2001yp,Acosta:2005ix,
Chekanov:2004kz,Abbiendi:2004pr,Abbiendi:2003cn,Buskulic:1995sw}.

For both signal and background we generate a number of events
equivalent to $1~\rm{fb}^{-1}$ of LHC data at $14$~TeV CM energy.  No
attempt is made at detector simulation through finite calorimeter
granularity, but we do impose a geometrical acceptance cut on jets of
$|\eta|<2.5$.

To illustrate the variety of approaches that can be taken with subjet
studies, we consider two complementary analyses: the first is an exclusive analysis with the $k_T$ algorithm~\cite{Catani:1993hr,Ellis:1993tq} and
searches for substructure in two jets, one for each neutralino
expected in an event. The other is an inclusive analysis, with the Cambridge/Aachen (C/A)
algorithm~\cite{Dokshitzer:1997in,Wobisch:1998wt}, which examines the
substructure of just the hardest jet. We find that the corresponding inclusive (exclusive) analysis with the $k_T$ (C/A) algorithm give similar results.

The $k_T$ algorithm defines distances
$d_{kl}\equiv\min{(p_{Tk}^{2},p_{Tl}^{2})}(\Delta
    R_{kl}^2/R^2),\,
  d_{kB}\equiv p_{Tk}^{2}$,
and sequentially merges the pair of objects $k,l$ with smallest $d_{kl}$,
unless there is a smaller $d_{kB}$, in which case $k$ becomes a jet.
The constant $R$ sets the angular reach of the jets.
Since the $k_T$ distance is just the relative transverse momentum
between objects, the mergers of interest for a decayed heavy-particle
tend to be the last ones. 
This was exploited in \cite{Butterworth:2002tt,Aad:2009wy,Butterworth:2007ke},
where a dimensionful cut was placed on the $d_{kl}$ scale of the
last merging in the jet, $d_1$, in order to preferentially select
boosted $W$ bosons over QCD jets, which, for a given mass, have
smaller $d_1$.

\begin{figure}[t]
  \centering
  \includegraphics[width=1.0\linewidth]{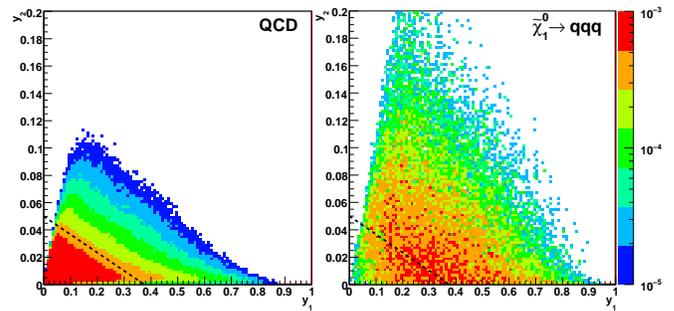}
  \caption{The $y$-values from the $k_T$ algorithm with $R=0.4$ for the last and
  next-to-last merging for QCD jets (left), and jets matched to neutralinos
  (right). Also shown is the proposed cut line.
  Both distributions are normalized to unity.} 
  \label{fig:y2D}
\end{figure}

However, our case differs from
\cite{Butterworth:2002tt,Aad:2009wy,Butterworth:2007ke} in two respects. First,
we are searching for an object of unknown mass, which means that we
should avoid biasing the search with a dimensionful substructure cut.
A good alternative is to cut on a dimensionless
variable normalised to the jet mass $m_j$, $y_i = d_i
R^2/m_j^2$. 
Secondly, the neutralino has a three-body decay, in contrast to a
W-boson's two-body decay. This suggests that one should cut on the
properties of the last \textit{two} mergers, i.e.\ on $y_1$ and
$y_2$.
Their distributions are illustrated in Fig.~\ref{fig:y2D}. 
The left panel shows QCD jets with $p_T>300$~GeV, while the right
panel shows jets matched to neutralinos for the SPS1a benchmark point,
the jet being within $\Delta R < 0.3$ of the neutralino with
$p_T>300$~GeV and mass 90~GeV~$<m_j<120$~GeV. Even after the hard cut
on jet $p_T$, there are clear differences between neutralino jets and
the QCD background.

For our analysis with the $k_T$ algorithm, we take $R=0.4$ and
use the following two cuts: i) at least four jets with
$p_T>300,300,100,100$~GeV and 
two of the high-$p_T$ jets ($p_T>300$~GeV) should have significant
substructure, $y_2>-0.13y_1 + 0.05$, consistent with fully or partially
contained neutralinos.
The choice of $R$ is a compromise between capturing sufficient signal,
favouring large $R$, and not smearing the mass peak with particles
from the underlying event, favouring small $R$~\cite{Dasgupta:2007wa}.
Cuts on a third and fourth jet are motivated by the expected
presence of two jets from the squark or gluino decay. We have verified that our simulation of QCD
events with additional jets is consistent with NLO
calculations in~\cite{Nagy:2001fj}.
The requirement of high $p_T$ should ensure high trigger efficiency,
as well as good collimation of the neutralino jets.

\begin{figure}[t]
  \centering
  \includegraphics[width=1.0\linewidth]{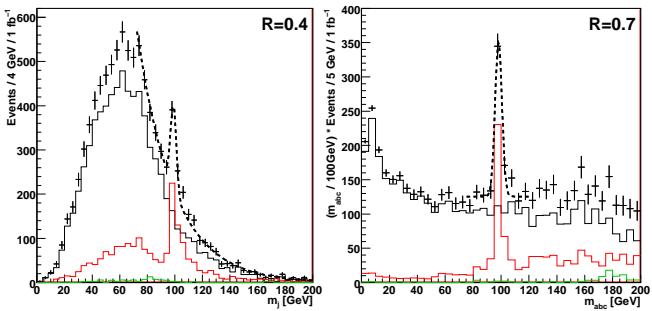}
  \caption{Jet mass distribution (points with error bars) for the $k_T$
  algorithm (left) and C/A algorithm (right) after all cuts. Also shown are the
  contributions from QCD (black), supersymmetric events (red) and other SM
  backgrounds (green).
  }
  \label{fig:mj}
\end{figure}

The mass distribution for the neutralino candidate jets after cuts is shown in
Fig.~\ref{fig:mj} (left). The $\tilde\chi_1^0$ is clearly visible as a perturbation on
the rapidly falling QCD background.
This analysis reconstructs 3.3\% of all neutralinos with $p_T>300$~GeV
(constituting 17.8\% of neutralinos produced) in a 20~GeV mass
window around the nominal mass; most reconstructed neutralinos
have a $p_T$ somewhat above the 300 GeV cut. For the QCD jets the
analysis accepts 0.023\% of all jets with $p_T>300$~GeV in the same
window.

Even though we have introduced only dimensionless jet substructure
cuts in the above analysis, the background distribution also has a
peak near the neutralino mass. This is a consequence of higher-order
perturbative effects \cite{Catani:1992ua,Almeida:2008yp} and the peak
position is determined by their interplay with $R$, the jet $p_T$ and
substructure cuts. Increasing $R$, which improves the signal
reconstruction efficiency, causes the background peak to move to
larger mass, potentially directly beneath the signal peak

To avoid this issue, we consider the C/A algorithm, which successively
recombines the pair of objects closest in $\Delta R_{kl}$, until all
objects are separated by more than $R$, at which point they are the
jets. Because the ordering of C/A mergers knows nothing about the
momentum scales involved, one cannot rely on the properties of the
last mergers to tag relevant substructure.

Instead we recurse through all mergers storing those subjets that are
sufficiently symmetric, having $z\equiv
\min(p_{Tk},p_{Tl})/(p_{Tk}+p_{Tl}) > z_{\min}$, ignoring in the
recursion the softer of the two subjets when $z<z_{\min}$.
From this we identify the two mergers that have the largest
\textsc{Jade}-type distance, $d^J_{kl} = p_{Tk} p_{Tl}\Delta R_{kl}^2$,
which is related to $m_{kl}^2$ if $k$ and $l$ are massless. If the
subjet resulting from the merger with smaller $d^J$ (labeling this
subjet ``bc'') is contained within the subjet from the other merger
(labeled ``abc''), i.e.\ the subjet ``abc'' is the result of
further mergers of ``bc'', then we consider ``abc'' to be a
neutralino candidate if $\mu \equiv m_{bc}/m_{abc} > \mu_{\min}$.
The cut on $z$ is used to avoid the soft splittings that dominate QCD
branching and that are largely responsible for producing the peak in
the mass distribution of QCD jets. The cut on the subjet mass ratio
$\mu$ ensures the presence of a three-body decay structure inside the
jet.

The full C/A-based analysis proceeds as follows: i) we use $R=0.7$ and
require at least four jets with $p_T>500,100,100,80$~GeV, and
$|\Delta\eta_{13}|,|\Delta\eta_{14}| < 1.5$, ii) the hardest jet is
taken to be a neutralino candidate if it passes the substructure cuts
with $z_{\min}=0.15$ and $\mu_{\min}=0.25$.
%
Relative to the $k_T$ analysis, the harder jet cuts help keep the
background under control despite requiring substructure in only one
jet.
%
%
For events that pass the cuts, we plot in Fig.~\ref{fig:mj}
(right) the distribution of $m_{abc}$, weighted with
$m_{abc}/100$~GeV. Expectations from QCD are that this distribution
should be rather flat for $m_{\min} \lesssim m_j \lesssim
p_TR\sqrt{z_{\min}}$, where $m_{\min}$ is some small value governed by
higher orders.
This is indeed what we observe, and for a range of choices of $R$
value and $p_T$ cut the signal is found to lie in this interval. The
analysis reconstructs 10.3\% of all neutralinos with $p_T>500$~GeV
(constituting 3.7\% of all neutralinos produced) in a 20~GeV mass
window around the nominal mass; for QCD jets 0.052\% of all jets with
$p_T>500$~GeV are accepted in this mass window.

\begin{figure}[t]
  \begin{center}
  \includegraphics[width=1.0\linewidth]{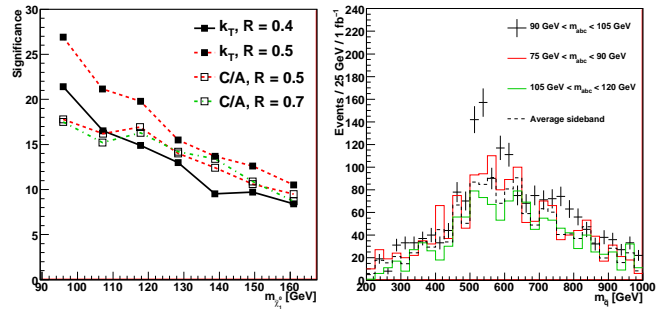}
  \caption{Estimated sensitivity for 1~fb$^{-1}$ as a function of
  $\tilde\chi_1^0$ mass for various choices of jet algorithm and size $R$
  (left), jet mass distribution for a squark search using the C/A
  algorithm with $R=0.5$ and background estimation by sidebands (right).}
  \label{fig:significance}
  \end{center}
\end{figure}

For both analyses we estimate the significance of the signal from the
number of signal and background events in a window of 20~GeV centered
on the peak, a choice consistent with the 7\% mass resolution seen in
ATLAS detector simulations~\cite{Aad:2009wy}.
This ignores the effect of the `looking-elsewere' problem, but should
demonstrate the potential of such a search. The results are shown in
Fig.~\ref{fig:significance} (left) for various neutralino masses along
the SPS1a benchmark line. Even in the highest-mass case studied, the
significance is well above the 5-$\sigma$ discovery `threshold' with
1~fb$^{-1}$ of statistics. In fact, the order of magnitude loss in
cross section over the parameter range is to a large extent
compensated by the lower QCD background at higher jet masses for the
$k_T$ analysis, and in the C/A case, jets from higher-mass squark and
gluino decays are more likely to pass our hard $p_T$ cut.

With evidence for a resonance peak, the next step is to estimate the
mass of the resonance. We fit the jet-mass distributions with a
background plus Gaussian signal distribution. The $k_T$ analysis
uses an exponential background in the interval $[80,200]$~GeV, while
the C/A analysis uses a uniform background in the interval
$[75,120]$~GeV. The results of this naive fit, which ignores the
experimental jet mass resolution, are shown in Table~\ref{tab:masses}
for the SPS1a benchmark point. Improvements on the systematic errors
inherent in this method are possible by calibrating the jet mass
against the known masses of the $W$ boson or top quark, for which a
reasonably clean measurement should be possible in events in which one
top quark decays leptonically. Improvements in mass measurement
are also possible through filtering of jets~\cite{Butterworth:2008iy}.

\begin{table}
  \centering
  \begin{tabular}{|l|c|c|c|c|}\hline
  Analysis & $m_{\tilde\chi_1^0}$ & $\chi^2/{\rm ndf}$ & $m_{\tilde q_R}$
           & $\chi^2/{\rm ndf}$  \\ \hline
  $k_T$, R=0.4 & ~$99.0\pm 0.5$~ & 1.43 & --- & --- \\ \hline
  C/A, R=0.5   & ~$98.2\pm 0.4$~ & 0.85 & $517.0\pm 2.5$ & 1.27 \\ \hline
  C/A, R=0.7   & ~$98.4\pm 0.6$~ & 1.46 & $526.6\pm 3.0$ & 1.20 \\\hline
  \end{tabular}
  \caption{Neutralino and squark mass fits for SPS1a. The nominal masses
  are $m_{\tilde\chi_1^0}=96.1$~GeV and $m_{\tilde q_R}=520$~GeV.
  }
  \label{tab:masses}
\end{table}

In Fig.~\ref{fig:significance} (right) we also demonstrate the
potential of our method for reconstructing the squark mass. By
selecting events from the C/A analysis in the signal band
90~GeV~$<m_{abc}<$~105~GeV and plotting both the mass of the
neutralino combined with the third jet and combined with the fourth
jet in the event, we arrive at the distribution in black, with a clear
peak around 520~GeV. The interpretation of the peak is checked by
plotting the sideband distributions, picking events from
75~GeV~$<m_{abc}<$~90~GeV (red) and 105~GeV~$<m_{abc}<$~120~GeV
(green). These show no sign of a peak. By subtracting the averaged
sidebands (dashed line), and fitting the remaining peak with a
Gaussian, we arrive at the squark mass estimates in
Table~\ref{tab:masses}.

The effects of pile-up, intrinsic resolution and granularity of the
detector will all have additional impact on the discovery of a
neutralino resonance and the measurement of its mass at the LHC, but
initial studies with realistic detector simulations indicate that the
efficiencies and resolutions assumed here are not
unreasonable~\cite{Aad:2009wy}. Further studies of the techniques
presented here are ongoing within the ATLAS collaboration~\cite{sky}.

In conclusion, we see that using sophisticated jet clustering
algorithms such as $k_T$ and C/A gives us the possibility of
discovering baryon-number violating decays of the type
$\tilde\chi_1^0\to qqq$, without the assumption of additonal features
such as hard leptons, and even when using only the substructure of the
hardest jet in the event.
%
%
We have further found that the neutralino mass can be measured to a
precision of a few GeV in these R-parity-violating scenarios, most
likely limited by the experimental jet mass resolution, and that one
can identify the squark resonance.
%
Realizing the potential outlined in the above analyses is a
challenge that merits experimental study.

This work was supported by the UK Science and Technology Facilities
Council (STFC). ARR whishes to thank the members of the Cambridge
Supersymmetry Working Group for many useful discussions. GPS thanks
Matteo Cacciari and Gregory Soyez for collaboration on FastJet
development and Pietro Slavich for helpful comments.


\bibliographystyle{h-physrev4-max4}

\bibliography{BERS}

\begin{thebibliography}{10}

\bibitem{Wess:1974tw}
J.~Wess and B.~Zumino,
\newblock Nucl. Phys. {\bf B70}, 39 (1974).

\bibitem{Fayet:1976cr}
P.~Fayet and S.~Ferrara,
\newblock Phys. Rept. {\bf 32}, 249 (1977).

\bibitem{Dimopoulos:1981zb}
S.~Dimopoulos and H.~Georgi,
\newblock Nucl. Phys. {\bf B193}, 150 (1981).

\bibitem{Nilles:1983ge}
H.~P. Nilles,
\newblock Phys. Rept. {\bf 110}, 1 (1984).

\bibitem{Haber:1984rc}
H.~E. Haber and G.~L. Kane,
\newblock Phys. Rept. {\bf 117}, 75 (1985).

\bibitem{:1999fr}
ATLAS Collaboration, I.~Hinchliffe {\em et~al.},
\newblock CERN-LHCC-99-15.

\bibitem{Allanach:2001xz}
B.~C. Allanach {\em et~al.},
\newblock JHEP {\bf 03}, 048 (2001), [hep-ph/0102173].

\bibitem{Allanach:2001if}
B.~C. Allanach {\em et~al.},
\newblock JHEP {\bf 09}, 021 (2001), [hep-ph/0106304].

\bibitem{Butterworth:2002tt}
J.~M. Butterworth, B.~E. Cox and J.~R. Forshaw,
\newblock Phys. Rev. {\bf D65}, 096014 (2002), [hep-ph/0201098].

\bibitem{Aad:2009wy}
ATLAS Collaboration, G.~Aad {\em et~al.},
\newblock arXiv:0901.0512.

\bibitem{Butterworth:2007ke}
J.~M. Butterworth, J.~R. Ellis and A.~R. Raklev,
\newblock JHEP {\bf 05}, 033 (2007), [hep-ph/0702150].

\bibitem{Butterworth:2008iy}
J.~M. Butterworth, A.~R. Davison, M.~Rubin and G.~P. Salam,
\newblock Phys. Rev. Lett. {\bf 100}, 242001 (2008), [arXiv:0802.2470].

\bibitem{Agashe:2006hk}
K.~Agashe {\em et~al.},
\newblock Phys. Rev. {\bf D77}, 015003 (2008), [hep-ph/0612015].

\bibitem{Brooijmans:2008se}
G.~H. Brooijmans {\em et~al.},
\newblock arXiv:0802.3715 [hep-ph].

\bibitem{Kaplan:2008ie}
D.~E. Kaplan, K.~Rehermann, M.~D. Schwartz and B.~Tweedie,
\newblock Phys. Rev. Lett. {\bf 101}, 142001 (2008), [arXiv:0806.0848].

\bibitem{Thaler:2008ju}
J.~Thaler and L.-T. Wang,
\newblock JHEP {\bf 07}, 092 (2008), [arXiv:0806.0023].

\bibitem{Almeida:2008yp}
L.~G. Almeida {\em et~al.},
\newblock arXiv:0807.0234.

\bibitem{Krohn:2009zg}
D.~Krohn, J.~Thaler and L.-T. Wang,
\newblock arXiv:0903.0392.

\bibitem{Ellis:2009su}
S.~D. Ellis, C.~K. Vermilion and J.~R. Walsh,
\newblock arXiv:0903.5081.

\bibitem{Allanach:2002nj}
B.~C. Allanach {\em et~al.},
\newblock Eur. Phys. J. {\bf C25}, 113 (2002), [hep-ph/0202233].

\bibitem{constraints}
We do not address the issue of constraints on this coupling, but note that our
  results are readily transferable to other couplings that are less restricted,
  such as $\lambda^{\prime\prime}_{212}$~\cite{Allanach:1999ic}.

\bibitem{Corcella:2000bw}
G.~Corcella {\em et~al.},
\newblock JHEP {\bf 01}, 010 (2001), [hep-ph/0011363].

\bibitem{Dreiner:1999qz}
H.~K. Dreiner, P.~Richardson and M.~H. Seymour,
\newblock JHEP {\bf 04}, 008 (2000), [hep-ph/9912407].

\bibitem{Moretti:2002eu}
S.~Moretti {\em et~al.},
\newblock JHEP {\bf 04}, 028 (2002), [hep-ph/0204123].

\bibitem{Corcella:2002jc}
G.~Corcella {\em et~al.},
\newblock hep-ph/0210213.

\bibitem{Pumplin:2002vw}
J.~Pumplin {\em et~al.},
\newblock JHEP {\bf 07}, 012 (2002), [hep-ph/0201195].

\bibitem{Butterworth:1996zw}
J.~M. Butterworth, J.~R. Forshaw and M.~H. Seymour,
\newblock Z. Phys. {\bf C72}, 637 (1996), [hep-ph/9601371].

\bibitem{jimmy}
The non-default parameter settings for {\sc Jimmy} are: PRSOF=0, PRRAD=1.8,
  PTJIM=4.9~GeV, JMUEO=1, corresponding to an ATLAS tune~\cite{Alekhin:2005dx}.

\bibitem{Cacciari:2005hq}
M.~Cacciari and G.~P. Salam,
\newblock Phys. Lett. {\bf B641}, 57 (2006), [hep-ph/0512210].

\bibitem{FastJetWeb}
M.~Cacciari, G.~P. Salam and G.~Soyez,
\newblock {FastJet},
\newblock \url{http://fastjet.fr/}.

\bibitem{Waugh:2006ip}
B.~M. Waugh {\em et~al.},
\newblock hep-ph/0605034.

\bibitem{Mangano:2002ea}
M.~L. Mangano {\em et~al.},
\newblock JHEP {\bf 07}, 001 (2003), [hep-ph/0206293].

\bibitem{Abazov:2001yp}
D0 Collaboration, V.~M. Abazov {\em et~al.},
\newblock Phys. Rev. {\bf D65}, 052008 (2002), [hep-ex/0108054].

\bibitem{Acosta:2005ix}
CDF Collaboration, D.~E. Acosta {\em et~al.},
\newblock Phys. Rev. {\bf D71}, 112002 (2005), [hep-ex/0505013].

\bibitem{Chekanov:2004kz}
ZEUS Collaboration, S.~Chekanov {\em et~al.},
\newblock Nucl. Phys. {\bf B700}, 3 (2004), [hep-ex/0405065].

\bibitem{Abbiendi:2004pr}
OPAL Collaboration, G.~Abbiendi {\em et~al.},
\newblock Eur. Phys. J. {\bf C37}, 25 (2004), [hep-ex/0404026].

\bibitem{Abbiendi:2003cn}
OPAL Collaboration, G.~Abbiendi {\em et~al.},
\newblock Eur. Phys. J. {\bf C31}, 307 (2003), [hep-ex/0301013].

\bibitem{Buskulic:1995sw}
ALEPH Collaboration, D.~Buskulic {\em et~al.},
\newblock Phys. Lett. {\bf B384}, 353 (1996).

\bibitem{Catani:1993hr}
S.~Catani, Y.~L. Dokshitzer, M.~H. Seymour and B.~R. Webber,
\newblock Nucl. Phys. {\bf B406}, 187 (1993).

\bibitem{Ellis:1993tq}
S.~D. Ellis and D.~E. Soper,
\newblock Phys. Rev. {\bf D48}, 3160 (1993), [hep-ph/9305266].

\bibitem{Dokshitzer:1997in}
Y.~L. Dokshitzer, G.~D. Leder, S.~Moretti and B.~R. Webber,
\newblock JHEP {\bf 08}, 001 (1997), [hep-ph/9707323].

\bibitem{Wobisch:1998wt}
M.~Wobisch and T.~Wengler,
\newblock hep-ph/9907280.

\bibitem{Dasgupta:2007wa}
M.~Dasgupta, L.~Magnea and G.~P. Salam,
\newblock JHEP {\bf 02}, 055 (2008), [0712.3014].

\bibitem{Nagy:2001fj}
Z.~Nagy,
\newblock Phys. Rev. Lett. {\bf 88}, 122003 (2002), [hep-ph/0110315].

\bibitem{Catani:1992ua}
S.~Catani, L.~Trentadue, G.~Turnock and B.~R. Webber,
\newblock Nucl. Phys. {\bf B407}, 3 (1993).

\bibitem{sky}
ATLAS Collaboration,
\newblock ATLAS-PHYS-PUB-2009-076.

\bibitem{Allanach:1999ic}
B.~C. Allanach, A.~Dedes and H.~K. Dreiner,
\newblock Phys. Rev. {\bf D60}, 075014 (1999), [hep-ph/9906209].

\bibitem{Alekhin:2005dx}
S.~Alekhin {\em et~al.},
\newblock hep-ph/0601012.

\end{thebibliography}

\end{document}